\documentstyle[12pt]{article}
\begin{document}

\begin{titlepage}

\begin{center}
{\bf ON ORIENTATION DEPENDENCE OF $N_2$ IONIZATION: THE VELOCITY
GAUGE VERSION OF MOLECULAR STRONG-FIELD APPROXIMATION}\\[30pt]

{\bf Vladimir I. Usachenko}{\it \footnote[1]{{\it Permanent
mailing address: Institute of Applied Laser Physics UzAS, Katartal
str. 28, Tashkent, 100135, Uzbekistan.
E-mail: alpi1990@uzsci.net.}}}$^{,2,3}$\\[10pt]

$^{1}${\it Institute of Applied Laser Physics UzAS, Katartal
str. 28, Tashkent, 100135, Uzbekistan,}\\[3pt]
$^{2}${\it Physics Department, National University of Uzbekistan,
Tashkent, 100174, Uzbekistan}\\[3pt]
$^{3}${\it Max-Born-Institute for Nonlinear Optics and Short-Pulse
Laser Spectroscopy,}\\
{\it Berlin, 12489, Germany}\\[30pt]
\end{center}

{\rm We reply to the Comment of T. K. Kjeldsen and L. B. Madsen
[e-print arXiv:physics/0508213] and acknowledge their criticism
related to imperfect composition of our model $3\sigma_g$
molecular state [V. I. Usachenko and S.-I. Chu, Phys. Rev. A {\bf
71}, 063410 (2005)]. However, we cannot agree with the authors
suggesting this critique also as an irrefutable argument and/or
evidence in support and justification of their opposite and {\it
incorrect} (viz. inconsistent with relevant experiment)
orientation dependence of $N_2$ ionization rate calculated within
the {\it velocity gauge} version of molecular strong-field
approximation. We disclose the actual reason of the contradiction
and demonstrate that appropriately composed $3\sigma_g$ state
(modified according to the Comment's critique) rather confirms the
alternative earler calculation [A. Jaron-Becker, A. Becker and F.
H. M. Faisal, Phys. Rev. A {\bf 69}, 023410 (2004)] suggesting
correct orientation dependence of $N_2$ ionization, contrary to
respective results of the Comment's authors applying {\it the
same} approach and procedure of $3\sigma_g$ composition.}

\end{titlepage}

\begin{center}
{\bf ON ORIENTATION DEPENDENCE OF $N_2$ IONIZATION: THE VELOCITY
GAUGE VERSION OF MOLECULAR STRONG-FIELD APPROXIMATION}\\[30pt]

{\bf Vladimir I. Usachenko}{\it \footnote[1]{{\it Permanent
mailing address: Institute of Applied Laser Physics UzAS, Katartal
str. 28, Tashkent, 100135, Uzbekistan.
E-mail: alpi1990@uzsci.net.}}}$^{,2,3}$\\[10pt]

$^{1}${\it Institute of Applied Laser Physics UzAS, Katartal
str. 28, Tashkent, 100135, Uzbekistan,}\\[3pt]
$^{2}${\it Physics Department, National University of Uzbekistan,
Tashkent, 100174, Uzbekistan}\\[3pt]
$^{3}${\it Max-Born-Institute for Nonlinear Optics and Short-Pulse
Laser Spectroscopy,}\\
{\it Berlin, 12489, Germany}\\[30pt]
\end{center}

\begin{enumerate}
\item  \noindent {\rm INTRODUCTION.}

\qquad {\rm The current {\it state-of-the-art} in strong-field
molecular ionization theory seems to be still far away from a
sufficiently clear insight due to numerous controversial results
obtained within different approaches and methods. As an
appropriate illustration to mention here is the problem of
strong-field ionization in laser-irradiated $N_2$ molecule, which
recently received a special consideration within both the {\it
velocity-gauge} (VG) [1-5] and the {\it length-gauge} (LG) [3, 6]
formulations of conventional {\it strong-field approximation}
(SFA) [7]. In particular, the orientation behavior of $N_2$
ionization rates was found in [3] to be quite different if
calculated in the VG version versus the LG version of the
molecular SFA approach and/or molecular tunneling theory [8].
Based on the same self-consistent Hartree-Fock (HF) numerical
procedure as previously proposed in [1, 2] for the $3\sigma_g$
{\it highest-occupied molecular orbital} (HOMO) composition, the
VG-SFA results [3] suggested a predominant $N_2$ ionization when
the molecular axis is perpendicular to the incident laser field
polarization (see also Fig. 2 in the Comment [4]). Thus, in
contrast to their LG-VGA results [3], the VG-SFA results of the
Comment's authors suggest a counterintuitive behavior for $N_2$
ionization, which contradict to both relevant experiment [9] and
alternative calculation [2]. Even though the Comment [4] applies
in fact {\it the same} VG-SFA approach and numerical (HF-based)
procedure of $3\sigma_g$ composition as earlier proposed in [1,
2], the Comment's authors carefully avoid of quoting the opposite
results of Ref.[2] and/or clarifying the reason of the
contradiction. They prefer instead to criticize our imperfect
composition of model $3\sigma_g$ molecular state [5] as
inappropriately symmetrized and oversimplified. Moreover, they
even suggest their criticism (being physically correct and
acceptable as such) as an indispensable explanatory reason of the
contradiction. In order to demonstrate the invalidity of such an
interpretation (as well as the related Comment's statements), we
have to identify the actual reason of the contradiction by means
of applying a more accurate molecular $3\sigma_g$ state
appropriately modified according to the Comment's critique.}\\

\item  \noindent {\rm  THE LCAO-MO METHOD AND ASSOCIATED
MOLECULAR WAVEFUNCTIONS OF $3\sigma_g$ STATE IN COORDINATE AND
MOMENTUM SPACE.}

\qquad {\rm According to the standard linear combination of atomic
orbitals (LCAO) and molecular orbitals (MO) method, the {\it
molecular orbitals} are the mathematical constructs used to model
the multi-electron molecular valence shells (each of a fixed
discrete binding energy} $\varepsilon _0^{\left( n\right)
}=-I_p^{(n)}$ {\rm and respective number }$N_e^{(n)}$ {\rm of
identical electrons, similar to {\it atomic orbitals} in atom).
For }$3\sigma _g$ {\rm HOMO of gerade and bonding symmetry in
}$N_2$ {\rm molecule under consideration, the corresponding
approximate two-centered single-electron molecular wavefunction
can be composed, for example, as an appropriate superposition of
scaled hydrogen-like }$2p_z$ {\rm and} $1s$ {\rm (or }$2s${\rm )
atomic orbitals:}
\begin{equation}
\Phi _{\left( 2p\right) 3\sigma _g}\left( {\bf r};{\bf R}_0\right)
=\sqrt \frac{N_e^{(3\sigma _g)}} {2 \left[ 1-S_{2p_z}^{(3\sigma
_g)}\left( R_0\right) \right] } \left[ \phi _{2p_z}^{(3\sigma
_g)}\left( {\bf r}-{\bf R}_0/2\right) -\phi _{2p_z}^{(3\sigma
_g)}\left( {\bf r}+{\bf R}_0/2\right) \right] \label{1}
\end{equation}
\begin{equation}
\Phi _{\left( 1s\right) 3\sigma _g}\left( {\bf r};{\bf R}_0\right)
=\sqrt \frac{N_e^{(3\sigma _g)}} {2 \left[ 1+S_{1s}^{(3\sigma
_g)}\left( R_0\right) \right] }\left[ \phi _{1s}^{(3\sigma
_g)}\left( {\bf r}-{\bf R}_0/2\right) +\phi _{1s}^{(3\sigma
_g)}\left( {\bf r}+{\bf R}_0/2\right) \right] \label{2}
\end{equation}
{\rm which are centered on each of the atomic cores and thus
separated by internuclear distance }$R_0${\rm . Here
$N_e^{(3\sigma _g)}=2$, whereas }
\begin{equation}
S_j^{(n)}\left( R_0\right) =\int d{\bf r}\phi _j^{(n)}\left( {\bf
r}+{\bf R} _0/2\right) \phi _j^{(n)}\left( {\bf r}-{\bf
R}_0/2\right) \label{3}
\end{equation}
{\rm denotes the respective atomic orbital overlap integral. }

\qquad {\rm Note that presently the ''$-$'' combination of $2p_z$
states in Eq.(1) is used, which does provide the even spatial
parity of the }$\left( 2p\right) 3\sigma _g$ {\rm HOMO. Thus, even
though the ''$+$'' combination of scaled hydrogen-like $2p_z$
atomic orbitals alone earlier used in the literature [5, 6] to
reproduce} $3\sigma _g$ {\rm HOMO in $N_2$ proved to be a
surprisingly well working in explaining relevant experiment [10,
11], we have to mention here the correct Comment's critical remark
that such combination doesn't provide the even ({\it gerade})
spatial parity and {\it bonding} symmetry of the }$\left(
2p\right) 3\sigma _g$ {\rm MO. Moreover, according to the MO-SFA
model [1], the oversimplified composition (1) would result in
destructive intramolecular interference (see also [1-2, 5]) and
accordingly high suppression in $N_2$ ionization (contrary to the
experiment [10, 11], which shows no suppression). To eliminate the
mentioned deficiency, in our present consideration a more accurate
composition of the }$3\sigma _g${ HOMO is applied taking into
account some admixture of a comparable contribution from atomic
$s$-states mostly required to provide a good agreement with
experiment (see also Sec.3 for details). Thus, in compliance with
the second Comment's critical remark, the }$3\sigma _g$ {\rm HOMO
is to be further approximated by the {\it coherent superposition}
of few different MOs corresponding to separate contributions from
atomic states of a specified orbital symmetry (viz., the scaled
hydrogen-like $1s$, $2s$ and $2p_z$ orbitals) under
consideration:}
\begin{equation}
\Phi _{\left( 1s2s2p\right) 3\sigma _g}\left( {\bf r};{\bf
R}_0\right) =A_{1s}\Phi _{\left( 1s\right) 3\sigma _g}\left( {\bf
r};{\bf R}_0\right) +A_{2s}\Phi _{\left( 2s\right) 3\sigma
_g}\left( {\bf r};{\bf R}_0\right) +A_{2p}\Phi _{\left( 2p\right)
3\sigma _g}\left( {\bf r};{\bf R}_0\right) \label{4}
\end{equation}
{\rm with the weight coefficients $A_j\ (j=1,2,3)$ being the
relative contributions ($\left| A_j\ \right| \leq 1$) from
respective atomic states and considered as variational parameters
to be found from the equation for minimum of respective molecular
binding energy, the value of which is put to be equal to the
experimental value.}

{\rm \qquad The model single-electron molecular wavefunctions
(1)-(2) allow for analytical representation of the Fourier
transform (or respective molecular state corresponding to a
definite value of final photoelectron momentum }${\bf p}$ {\rm in
momentum space) for each of one-electron two-centered single
molecular orbitals contributing to the }$3\sigma _g$ {\rm HOMO in
}$N_2$ {\rm under consideration:}
\begin{equation}
\widetilde{\Phi }_{\left( 2p\right) 3\sigma _g}\left( {\bf
p}_N{\bf ,R}_0\right) = - \sqrt {N_e^{(3\sigma _g)}} C\left(
\kappa _n\right) \frac{2^5\kappa _n^{7/2}p_N\cos \left( \theta
_{{\bf p} }\right) }{\pi \sqrt{2}\left( p_N^2+\kappa _n^2\right)
^3}\frac{\sin \left[ \left( {\bf p}_N{\bf \cdot R}_0\right)
/2\right] }{\sqrt{2\left[ 1-S_{2p_z}\left( R_0\right) \right] }}
\label{5}
\end{equation}
\begin{equation}
\widetilde{\Phi }_{\left( 2s\right) 3\sigma _g}\left( {\bf
p}_N{\bf ,R}_0\right) = \sqrt {N_e^{(3\sigma _g)}} C\left( \kappa
_n\right) \frac{2^4\kappa _n^{5/2}\left( p_n^2-\kappa _N^2\right)
}{ \pi \sqrt{2}\left( p_N^2+\kappa _n^2\right) ^3}\frac{\cos
\left[ \left( {\bf p}_N{\bf \cdot R}_0\right) /2\right]
}{\sqrt{2\left[ 1+S_{2s}\left( R_0\right) \right] }}  \label{6}
\end{equation}
\begin{equation}
\widetilde{\Phi }_{\left( 1s\right) 3\sigma _g}\left( {\bf
p}_N{\bf ,R}_0\right) = \sqrt {N_e^{(3\sigma _g)}} C\left( \kappa
_n\right) \frac{2^{5/2}\kappa _n^{5/2}}{\pi \left( p_N^2+\kappa
_n^2\right) ^2}\frac{\cos \left[ \left( {\bf p}_N{\bf \cdot
R}_0\right) /2\right] }{\sqrt{2\left[ 1+S_{1s}\left( R_0\right)
\right] }} \label{7}
\end{equation}
{\rm with the polar angle }$\theta _{{\bf p}}$ {\rm of
photoelectron emission relative to internuclear molecular axis
}$\left( \cos \left( \theta _{{\bf p}}\right) =\left( {\bf p}\cdot
{\bf R}_0\right) /pR_0\right)${\rm . Here }$\varepsilon _0^{\left(
n\right) }=-\kappa _n^2/2=-I_p^{(n)}=-\left( Z_j^{(n)}/a_j\right)
^2/2$ {\rm is the binding energy of }$n${\rm -th initial molecular
discrete state under consideration (viz., } $I_p^{\left( 3\sigma
_g \right)}\left( N_2\right) \approx 15.58$ $eV${\rm ) and
}$Z_j^{(n)} ${\rm is the effective charge corresponding to ''{\it
effective}'' long-range Coulomb model binding potential of
respective residual molecular ion, while }$ a_j=ja_0$ {\rm is
}$j${\rm -th Bohr orbital radius of respective contributing atomic
orbital. The analytical expressions for respective atomic orbital
overlap integrals (3) are presented in [5], moreover, the
Coulomb-Volkov correction factor }$ C\left( \kappa _n\right)
=\left( 2\kappa _nI_p^{(n)}/E\right) ^{\kappa _n^{-1}}${\rm \ is
also introduced to matrix elements (5)-(7) to incorporate the
long-range Coulomb electron-molecular ion interaction in the final
continuum states due to the VG version of SFA applied [1]}.\\

\item  \noindent {\rm DISCUSSION.}

\qquad {\rm Fig.1 demonstrates the angular dependence of molecular
wavefunction (4) in momentum space presented for different
$3\sigma _g$ compositions, either with or without taking a
comparable contribution from atomic $s$-states into account. Let's
note first that the angular dependence corresponding to
contribution from }$2p_z$ {\rm states alone (Fig.1a) is
considerably prolate along the internuclear axis since the $2p_z$
states are predominantly ionized along the internuclear axis owing
to the presence of the geometrical factor $\cos \theta _{{\bf p}
}$ in Eq.(5) (cf. also with Eq.(20) and respective Fig.1a in [5]).
This, particularly, illustrates that the spatial (gerade or
ungerade) symmetry of respective }$\left( 2p\right) 3\sigma _g$
{\rm MO doesn't essentially affect the angular dependence (viz.,
{\it nodal plane} along the molecular axis) in momentum space.
Thus, irrespectively of the sign in the combination of $2p_z$
atomic states, the respective }$\left( 2p\right) 3\sigma _g$ {\rm
MO suggests a predominant photoelectron emission along the
internuclear axis in momentum space. The sign in the combination
mostly alters the absolute value of local maxima in respective
angular dependence due to different ({\it sine} or {\it cosine})
trigonometric factor arising from intramolecular interference of
contributions of ionization from two separate atomic centers [1].
In particular, the $\sin \left[ \left( {\bf p}_N{\bf \cdot
R}_0\right) /2\right]$ factor in Eq.(5) corresponding to the
''$-$'' sign results in {\it destructive} intramolecular
interference and a suppressed molecular ionization (e.g. in $O_2$
molecule relative to $Xe$ atom, see also [1, 2, 5], for details).}

\qquad {\rm Accordingly, if the currently used ''$-$'' combination
of $2p_z$ states is supposed to contribute to the $3\sigma _g$ MO,
then a coherent contribution from other ($1s$ and/or $2s$) atomic
states should be also taken into equally detailed consideration as
described by Eq.(4). The respective Figures 1c-1d (versus Fig.1a)
demonstrate that the contribution from $s$-states to the
}$\left(1s2p\right) 3\sigma _g$ {\rm and }$\left(1s2s2p\right)
3\sigma _g$ {\rm MO dominates in momentum space at low
photoelectron momenta, in a qualitative accordance with stated in
the Comment. Unfortunately, the Reply's restrictions don't allow
us to provide the reader with an extensive and sufficiently
detailed information about the respective molecular ionization
rates, it will be given elsewhere in our further related
publications. So far we only have to stress that the corresponding
relative (weight) coefficients $A_j$ at $1s$, $2s$ and $2p_z$
states under consideration proved to have comparable values,
although not necessarily of the same sign. Namely, the $3\sigma
_g$ compositions with $A_{1s}$ and $A_{2p}$ (as well as $A_{1s}$
and $A_{2s}$) {\it of opposite signs} only proved to provide no
suppression observed in experiment [10, 11] for $N_2$ ionization
as compared to ionization of counterpart atomic $Ar$ of nearly
identical ionization potential $I_p^{(Ar)}\approx 15.75$ $eV$. In
particular , the total ionization rates calculated for $A_{1s}$
and $A_{2p}$ of opposite signs in $\left( 1s2p\right) 3\sigma _g$
and $\left(1s2s2p\right) 3\sigma _g$ MO proved to demonstrate no
suppression (and even some enhancement) in $N_2$ ionization
relative to }$Ar$ {\rm well consistent with relevant experiment
[10, 11]. Our Figs.1c-1d also suggest that $\left( 1s2p\right)
3\sigma _g$ and $\left(1s2s2p\right) 3\sigma _g$ {\rm MO
corresponding to the coefficients }$A_{1s}$ and $A_{2p}$ of
opposite signs are always noticeably prolate in momentum space
along the internuclear axis, although to a considerably less
extent as compared to the }$(2p)3\sigma _g $ {\rm composed of pure
$2p_z$ states alone.}

\qquad {\rm Anyhow, our Figs.1c-1d clearly illustrate that a more
accurate consideration doesn't necessarily result in the $3\sigma
_g$ HOMO to predominate in momentum space along the direction
perpendicular to the internuclear axis (as suggested, e.g., by
Fig.4e of the first reference in [3] and/or Fig.1f in the Comment
[4]). Since the generalized Bessel function are commonly maximized
when photoelectron momentum }${\bf p}$ {\rm is parallel to the
laser field polarization }${\bf e}${\rm , the ionization from
}$\left( 1s2p\right) 3\sigma _g$ and $\left(1s2s2p\right) 3\sigma
_g$ {\rm presented in our Fig.1 is expected to be predominant when
the} $N_2$ {\rm molecular axis is aligned along the laser field
}(${\bf R}_{{0}}||{\bf e}$){\rm . This difference thus invalidates
the Comment's statement that incorrect (viz., inconsistent with
experiment) orientation dependence of $N_2$ ionization rate the
authors also found in [3] is just explained by the VG version of
molecular SFA they applied. Our Fig.2 well illustrates the above
conclusion and displays the orientation dependence of total }$N_2$
{\rm ionization rates (see Eq.(13) in [5]) corresponding to
different }$3\sigma _g$ {\rm compositions. As expected, the
compositions of }$3\sigma _g $ {\rm HOMO taking a comparable
contribution from $p$-states into account always suggest a
predominant ionization of }$N_2$ {\rm for parallel orientation of
internuclear axis relative to the laser field polarization.
Whereas, the composition of }$3\sigma _g$ {\rm accounting for a
contribution from pure $s$-states alone is well seen to only give
a predominant ionization of }$N_2$ {\rm for perpendicular
orientation of internuclear axis as suggested in the Comment. Such
a behavior may thus arise only due to the contributing $1s$ or
$2s$ states, which are {\it spherically symmetric} and equally
well ionized along any spatial direction. Therefore, the form of
respective nodal plane relative to the molecular axis suggested by
Eqs.(6) and (7) is solely determined by the {\it cosine}
trigonometric factor }$\cos \left[ \left( {\bf p}_N{\bf \cdot
R}_0\right) /2\right]${\rm , which alone suggests the angular
dependence to be {\it oblate} along the internuclear axis in
momentum space (Fig.1b). The resulting orientation dependence with
maximum ionization for the perpendicular orientation of molecular
axis is thus quite naturally expected for ionization of $H_2$ and
$H_2^{+}$ (with a single} $\left(1s\right)1\sigma _g$ {\rm MO
composed of pure $s$-states). For $H_2$, such orientation
dependence was recently revealed [12] using a tunnelling model
with a Heitler–-London type valence bond wavefunction.}

\qquad {\rm To summarize, our argumentation based on a more
accurate consideration of }$3\sigma _g$ {\rm HOMO (4) is also
qualitatively well consistent with alternative calculation [2]
using {\it the same} VG-SFA approach as applied by the Comment's
authors [3, 4]. Meantime, we also have to stress that opposite
orientation dependence of $N_2$ ionization presented in Fig.2 of
the Comment [4] is based on {\it the same} numerical HF-based
procedure of $3\sigma_g$ HOMO composition as applied in [2] under
precisely {\it the same} situation (cf. with respective Fig.7b in
[2]). Therefore, we believe that orientation dependence of }$N_2$
{\rm ionization rate presented in our Fig.2 is correct and its
consistence with relevant experiment [9] not just an accidental.
By means of direct comparison of our Fig.1c (and/or Fig.1d) with
respective Fig.4e presented in [3], one can identify that opposite
behavior of $N_2$ ionization suggested in [4] (as well as earlier
in [3]) stems from a different angular dependence of $3\sigma_g$
molecular state in momentum space. Our Fig.1 also suggests that
such difference may only arise owing to relative contribution from
$s$-states, which seems to be somehow {\it overestimated}
(compared to $p$-states) within the particular (12s7p)/[6s4p]
basis chosen in the Comment for numerical HF-based procedure of
$3\sigma_g$ composition in coordinate space. This, particularly,
becomes especially evident from Fig.3 displaying the correct
orbital shape for the $3\sigma _g$ wavefunction in coordinate
space presented in Ref.[13] (see Fig.12 therein), which obviously
differs by a remarkably prominent domination of $p$-states from
the resulting $3\sigma _g$ state of substantially different shape
and detailed structure suggested in [4] (cf. respective Fig.1c
therein). Meantime, the numerical procedure applied in [3, 4] for
$3\sigma _g$ composition doesn't seem to be sufficiently
transparent with respect to the extent of such relative
contribution, which is thus hardly amenable to reliable analysis
and/or available for direct comparison [14]. Note also that a
composition overestimating the relative contribution from
$s$-states in coordinate space is not suitable to $3\sigma_g$
state in $N_2$, which is known to have a different and more
complex structure well {\it dominated} by $p$-states [15, 16].}

\qquad {\rm Finally, we also have to clarify the main message of a
different (although, somewhat related) research [17] the Comment's
authors quite inappropriately and very ambiguously invoked in
support of their related statement suggesting the LG version to be
superior to the VG version of the SFA under discussion. Recall
that the cited similar conclusion of Ref.[17] (see also Fig.1
therein to make certain) applies solely to ionization of {\it
atomic} states of only {\it odd} spatial parity, such as
$p$-states, which are not symmetric with respect to the inversion
of all electron coordinates to opposite direction. Whereas, for
the {\it molecular} $3\sigma_g$ state (which is known as surely
{\it gerade}, i.e. of {\it even} spatial parity), the VG and LG
versions of SFA, according to [17], are expected to give {\it the
same} results, contrary to respective VG-SFA findings of Ref.[3]
and the context of main related statement of the Comment [4].}\\

\item  {\rm \noindent CONCLUDING REMARKS.}

\qquad {\rm The criticism of the Comment [4] related to our
previous imperfect composition of the $3\sigma_g$ molecular
wavefunction [5] was assessed to be physically correct as such,
however not affecting essentially our related final results and
conclusions based on the velocity gauge (VG) version of molecular
strong-field approximation (SFA). In particular, when applying an
appropriately modified $3\sigma_g$ wavefunction, we demonstrate
that orientation dependence of $N_2$ ionization is to be correct
(viz., consistent with experiment) and thus the Comment's critique
cannot be accepted as an irrefutable argument and indispensable
evidence in support of the authors' opposite respective VG-SFA
findings [3, 4]. This, particularly, invalidates the Comment's
main statement about unavailability of the SFA in its VG
formulation for adequate description of molecular strong-field
ionization when an accurate molecular orbital is applied. Our
conclusion is also well consistent with recent Erratum [18]
suggesting in fact analogous corrections to analytical expressions
for molecular ionization rate previously numerically calculated in
[1, 2] for diatomic molecule having a HOMO of gerade symmetry.
Those corrections were also confirmed in [18] as not affecting
their correct final results and conclusions discussed elsewhere
earlier including the interference suppression of molecular
ionization and orientation dependence of $N_2$ ionization [1, 2],
which proved to be quite opposite to respective VG-SFA result of
the Comment applying {\it the same} approach and pure numerical
HF-based procedure of $3\sigma_g$ composition.}\\

\item  \noindent {\rm ACKNOWLEDGMENTS.}

\qquad {\rm This work was supported by Center of Sciences and
Technology of Uzbekistan (Project No. }$\Phi ${\rm -2.1.44). The
support from DAAD (Deutscher Akademischer Austauschdienst) and
B-Division of Max-Born-Institute of Nonlinear Optics and
Short-Pulse Laser Spectroscopy (Berlin, Germany) is also
gratefully acknowledged.}

\qquad {\rm The research described in this presentation was made
also possible in part by financial support from the U.S. Civilian
Research and Development Foundation (CRDF) for the Independent
States of the Former Soviet Union.}
\end{enumerate}

\begin{center}
{\rm {\bf FIGURE\ CAPTIONS}}\\[30pt]
\end{center}

\begin{enumerate}
\item {\rm Fig.1 (Color online). The squared module $\left|
\widetilde{\Phi }_{3\sigma _g}\left({\bf p}_N{\bf ,R}_0\right)
\right| ^2$ of $3\sigma _g$ HOMO wavefunction in momentum space
(the Fourier transform) plotted vs the angle $\theta_{{\bf p}}$ of
photoelectron emission (relative to the internuclear molecular
axis lined up in the horizontal direction) and calculated for four
different $3\sigma _g$ compositions: (a) $ \left( 1s\right)
3\sigma_g$ MO composed from $1s$ states alone, (b) $\left(
2p\right) 3\sigma_g$ MO composed from $2p_z$ states alone, (c)
$\left( 1s2p\right) 3\sigma _g$ MO composed from $1s$ and $2p_z$
states, (d) $\left( 1s2s2p\right) 3\sigma_g$ MO composed from
$1s$, $2s$ and $2p_z$ states. These angular dependencies are all
presented for photoelectrons only emitted along the polarization
of {\it Ti:sapphire} incident laser field of the wavelength
$\lambda =800$ $nm$ ($\hbar \omega =1.55\ eV$) and fixed intensity
$I=2\cdot 10^{14}W/cm^2$ due to absorption of different number $N$
of laser photons beginning from minimum one $N_0=18$ required for
ionization of $N_2$.}

\item {\rm Fig.2 (Color online). The total $N_2$ ionization rates
corresponding to $3\sigma _g$ compositions presented in Fig.1 and
plotted vs the angle $\Theta $ of the internuclear axis
orientation relative to the polarization of {\it Ti:sapphire}
incident laser field of the wavelength $\lambda =800$ $nm$ ($\hbar
\omega =1.55\ eV$) and fixed intensity $I=2\cdot 10^{14}W/cm^2$.
The ionization rates are all normalized to respective ones at the
angle $\Theta=0$ and presented for four different $3\sigma _g$
compositions: $\left( 1s\right) 3\sigma _g$ MO composed from $1s$
states alone (solid line); $\left( 2p\right) 3\sigma _g$ MO
composed from $2p_z$ states alone (dashed-dotted line); $\left(
1s2p\right) 3\sigma _g$ MO composed from $1s$ and $2p_z$ states
(short dashed line), $\left( 1s2s2p\right) 3\sigma _g$ MO composed
from $1s$, $2s$ and $2p_z$ states (long dashed line). This figure
is to be compared with Fig.2a of the Comment and Fig.2a of
relevant experiment [9].}

\item {\rm Fig.3 (Color online). The orbital shape of the correct
$3\sigma _g$ HOMO in $N_2$ molecule imaged by recording the high
harmonic spectra and presented in [13] (see Fig.12 therein and
related explanation). Note that this is not the orbital density,
but a wavefunction to within a global phase. This figure is to be
compared with respective Fig.1c of the Comment displaying the
alternative $3\sigma _g$ HOMO numerically composed within
particular $(12s7p)/[6s4p]$ basis set.}

\end{enumerate}


\begin{thebibliography}{99}

\bibitem{[1]}  {\rm J. Muth-B\"{o}hm, A. Becker, and F. H. M. Faisal,
Phys. Rev. Lett. {\bf 85}, 2280 (2000).}

\bibitem{[2]} {\rm A. Jaron-Becker, A. Becker and F. H. M. Faisal,
Phys. Rev. A {\bf 69}, 023410 (2004).}

\bibitem{[3]}  {\rm T. K. Kjeldsen and L. B. Madsen, J. Phys. B {\bf 37},
2033 (2004); Phys. Rev. A {\bf 71}, 023411 (2005).}

\bibitem{[4]}  {\rm T. K. Kjeldsen and L. B. Madsen, e-print arXiv:physics/0508213.}

\bibitem{[5]}  {\rm V. I. Usachenko and S.-I Chu, Phys. Rev. A {\bf 71},
063410 (2005).}

\bibitem{[6]}  {\rm K. Mishima, M. Hayashi, J. Yi, S. H. Lin, H. L. Selzle
and E. W. Schlag, Phys. Rev. A {\bf 66}, 033401 (2002).}

\bibitem{[7]}  {\rm H. R. Reiss, Prog. Quantum Electron. {\bf 16}, 1
(1992); H. R. Reiss, Phys. Rev. A {\bf 42}, 1476 (1990).}

\bibitem{[8]}  {\rm X. M. Tong, Z. X. Zhao and C. D. Lin, Phys. Rev. A
{\bf 66}, 033402 (2002); Z. X. Zhao, X. M. Tong, and C. D. Lin,
Phys. Rev. A {\bf 67}, 043404 (2003).}

\bibitem{[9]}  {\rm I. V. Litvinyuk, et al., Phys. Rev. Lett. {\bf 90},
233003 (2003).}

\bibitem{[10]}  {\rm C. Guo, M. Li, J. P. Nibarger and G. N. Gibson,
Phys. Rev. A {\bf 58}, R4271 (1998).}

\bibitem{[11]}  {\rm E. Wells, M. J. DeWitt and R. R. Jones, Phys.
Rev. A {\bf 66}, 013409 (2002); M. J. DeWitt, E. Wells and R. R.
Jones, Phys. Rev. Lett. {\bf 87}, 153001 (2001).}

\bibitem{[12]}  {\rm A. A. Kudrin and V. P. Krainov, Laser Phys.,
{\bf 13}, 1024 (2003).}

\bibitem{[13]}  {\rm A. Scrinzi, M. Yu. Ivanov, R. Kienberger
and D. M. Villeneuve, J. Phys. B {\bf 39}, R1 (2006).}

\bibitem{[14]}  {\it Unfortunately, the authors of Ref.[3] ignored
our urgent request to provide their Comment [4] with a reliable
quantitative information about the extent of these relative
contributions under discussion.}

\bibitem{[15]}  {\rm A. C. Wahl, Science {\bf 151}, 961 (1966).}

\bibitem{[16]}  {\rm B. Zimmermann, M. Lein and J. M. Rost, Phys. Rev.A
{\bf 71}, 033401 (2005).}

\bibitem{[17]}  {\rm D. Bauer, D. B. Milosevic, and W. Becker, Phys.
Rev. A {\bf 72}, 023415 (2005).}

\bibitem{[18]}  {\rm J. Muth-B\"{o}hm, A. Becker, and F. H. M. Faisal,
Phys. Rev. Lett. {\bf 96}, 039902(E) (2006).}

\newpage
\end{thebibliography}
\end{document}